\documentclass[%
 reprint,
superscriptaddress,
 amsmath,amssymb,
 aps,
prb,
]{revtex4-2}

\usepackage{graphicx}% Include figure files
\usepackage{dcolumn}% Align table columns on decimal point
\usepackage{bm}% bold math
\usepackage{siunitx}

\usepackage{multirow} %% multirow tables
\usepackage{chemformula}
\mathchardef\mhyphen="2D

\begin{document}

\preprint{APS/123-QED}

\title{Low temperature magnetism of KAgF$_3$}% Force line breaks with \\
\author{John M. Wilkinson}%
\affiliation{%
 University of Oxford, Department of Physics, Parks Rd, Oxford OX1 3PU, UK}
 \affiliation{STFC-ISIS Neutron and Muon Source, Harwell Campus, Oxfordshire, OX11 0QX, UK}%Lines break automatically or can be forced with \\
\author{Stephen J. Blundell}%
 \email{stephen.blundell@physics.ox.ac.uk}
\affiliation{%
 University of Oxford, Department of Physics, Parks Rd, Oxford OX1 3PU, UK}%


\author{Sebastian Biesenkamp}
\author{Markus Braden}%
 \email{braden@ph2.uni-koeln.de}
 \affiliation{
 II. Physikalisches Institut, Universität zu Köln, Zülpicher Str. 77, D-50937 Köln, Germany\\
}%
\author{Kacper Koteras}
\author{Wojciech Grochala}%
 \email{w.grochala@cent.uw.edu.pl}
 \affiliation{%
 University of Warsaw, Center of New Technologies, Żwirki i Wigury 93, 02-089 Warsaw, Poland\\
}%

\author{Paolo Barone}
 \affiliation{%
Superconducting and Other Innovative Materials and Devices Institute (SPIN), Consiglio Nazionale delle Ricerche, Area della Ricerca di Tor Vergata, Via del Fosso del Cavaliere 100, I-00133 Rome, Italy\\
}%

\author{Jos{\'e} Lorenzana}%
 \email{jose.lorenzana@cnr.it}
 \affiliation{%
Institute for Complex Systems (ISC), Consiglio Nazionale delle Ricerche, Dipartimento di Fisica,
Università di Roma “La Sapienza”, 00185 Rome, Italy\\
}%

\author{Zoran Mazej}
\author{Ga{\v s}per Tav{\v c}ar}%
 \affiliation{%
Jožef Stefan Institute, Department of Inorganic Chemistry and Technology, Jamova cesta 39, 1000 Ljubljana, Slowenia\\
}%

\date{\today}% It is always \today, today,

\begin{abstract}
 KAgF$_3$ is a quasi one-dimensional quantum antiferromagnet hosting a series of intriguing structural and magnetic transitions. Here we use powder neutron diffraction,  $\mu$SR spectroscopy, and Density Functional Theory calculations to elucidate the low temperature magnetic phases. Below $T_{N1}=29$K we find that the material orders as an A-type antiferromagnet with an ordered moment of 0.47~$\mu_{\rm B}$. Both neutrons and muons provide evidence for an intermediate phase at temperatures $T_{N1}<T<T_{N2}$ with $T_{N2}\approx 66$ K from a previous magnetometry study. However, the evidence is at the limit of detection and its nature remains an open problem.
%\begin{description}
%\end{description}
\end{abstract}

\maketitle

%\tableofcontents

\section{\label{sec:intro} Introduction}
Spin-1/2 systems show a profusion of fascinating phenomena due to strong quantum fluctuations, as the 
quantum effects in many-body systems increase when the mass and angular momentum of their constituents decrease. 
For example, light He remains liquid at zero temperature due to quantum fluctuations. By analogy, Anderson \cite{Anderson1973} proposed that 
spin-1/2 systems may form spin-liquid states at low temperatures, which lack long-range magnetic order. While at high dimension, the existence of spin-liquids is controversial \cite{Sorella2012}, they certainly form in quasi-one-dimensional magnetic systems \cite{Dutton2012, Savary2016}. The elementary excitations called spinons have been observed with a variety of probes \cite{Suzuura1996,Lorenzana1997,Schlappa2018,Enderle2010,Lake2013,Mourigal2013}. Spinons behave as Fermions and can be visualized as domain-walls of the local antiferromagnetic order. Although physical realizations of spin-1/2 systems are good correlated insulators, they can conduct heat as one-dimensional metals due to the fermionic properties of spinons \cite{Hess2019}. 
Additionally, due to the Fermionic nature of excitations, spin-1/2 systems may support a 
spin-Peierls instability \cite{Giamarchi2004}, the magnetic analogue of the Peierls instability in the one-dimensional electron gas.

 Spin-1/2 realizations in condensed matter are represented mostly by Ti(III) or V(IV) compounds with the $d^1$ electronic configuration of the metal cation, and by Cu(II) systems with a 
 $d^9$ electron configuration. A great diversity of the Cu(II) systems have been described 
 in the literature with remarkable properties. For example,
 two-dimensional layered systems include the celebrated parent compounds \cite{Kastner1998} 
 of high-$T_\mathrm{c}$ superconductors. Their strong quantum fluctuations (unlike higher-spin systems) have been recently exposed in resonant inelastic x-ray scattering studies (RIXS) \cite{Betto2021,Martinelli2022}.  
 \ch{CuGeO3} is a nice example of inorganic spin-Peierls system with a structural transition clearly linked to the magnetism \cite{Hase1993a,Braden1996a}. \ch{Sr2CuO3} is an almost perfect realization of the one-dimensional  Heisenberg model, showing an excellent example of a spinon spectra in optical properties \cite{Suzuura1996,Lorenzana1997} and multispinon excitations in RIXS \cite{Schlappa2018}. A gigantic spinon mean-free path has also been deduced from thermal transport\cite{Hess2019}.
  CuO is a quasi one-dimensional spin-1/2 system with much lower symmetry than \ch{Sr2CuO3}. 
  On lowering the temperature, it enters first into an incommensurate magnetic spiral phase 
  and then into a commensurate antiferromagnetic phase. The incommensurate phase is also 
  multiferroic \cite{Kimura2008}. The rich phenomenology of CuO can be understood from 
  the intrinsic frustration built into the structure\cite{Giovannetti2011a,Hellsvik2014}. 
 
The analogous heavier Ag(II) congeners are much less researched \cite{Grochala2001} but their similarities to cuprates \cite{Gawraczynski2019,Bachar2022,Piombo2022} calls for an exploration of this family searching for analogous rich physics. KAgF\textsubscript{3} is one of such systems which shows  intriguing structural and magnetic phenomena \cite{Kurzydlowski2013}.

\begin{figure}
    \centering
    \includegraphics[width=0.8\columnwidth]{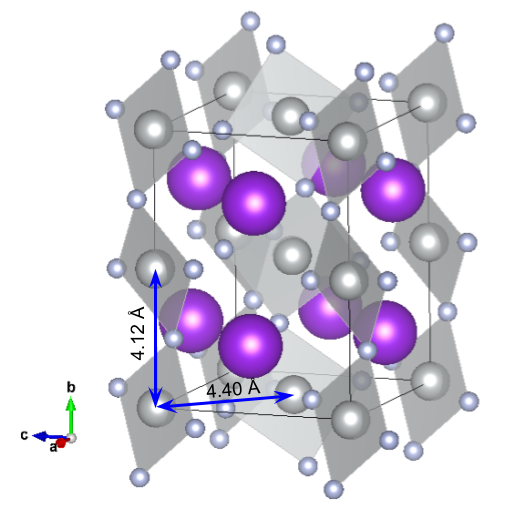}
    \caption{ Structure of \ch{KAgF3}. Spheres represent (in decreasing size order): K, Ag and F. Grey plaquettes highlight short Ag-F bonds. Thin black lines indicate the orthorhombic unit cell.}
    \label{fig:struct}
\end{figure}

This compound crystallizes in an orthorhombic unit cell (Figure~\ref{fig:struct}) which may be viewed as a distorted cubic perovskite structure. Silver-silver distances are significantly shorter along the ${b}$ axis, as indicated in the figure. The gray plaquettes indicate short Ag--F bonds of elongated \ch{AgF6} octahedra, with the long axis approximately perpendicular to the plaquette. The short Ag--Ag bond (4.12~\AA) along the  ${b}$ axis defines  kinked chains which host very strong antiferromagnetic superexchange 
with the coupling constant of the order of $\approx 100$~meV from susceptibility measurements \cite{Kurzydlowski2013} or even larger according to DFT computations (see Refs.~\cite{Kurzydowski2017,Kurzydowski2017b,Zhang2011} and Supplementary Information). Simultaneously, the interchain interactions are much weaker and of the order of few meV. One can visualize the plaquettes as hosting Ag orbitals with approximately $d_{x^2-y^2}$-symmetry, with lobes 
pointing towards the fluorines. These mix with $p$-orbitals in F along the chain, providing a robust path for superexchange \cite{Gawraczynski2019} as in \ch{AgF2}.  

A previous study \cite{Kurzydlowski2013} documented a structural transition near 
$T=235$~K, accompanied by an intriguing drop of the susceptibility on lowering the temperature, which is reminiscent of a spin-Peiers transition.  Magnetic order appears at lower temperatures with two  magnetic transitions at $T_{N2}=66$~K and $T_{N1}=35$~K, whose nature has not been explored 
in depth so far.
DFT studies suggest predominant antiferromagnetic (AF) interactions \cite{Zhang2011,Kurzydlowski2013}, thus we tentatively identify 
$T_{N1}$ and $T_{N2}$ as N\'eel temperatures using the notation that is common in CuO.

The present work aims at elucidating the low temperature magnetic phases of \ch{KAgF3} using 
 $\mu$SR spectroscopy, powder neutron diffraction, and Density Functional Theory calculations. 
Neutron scattering experiments were performed at ILL, and $\mu$SR experiments were performed at ISIS and PSI. The continuous muon source at PSI 
is particularly suited for detailed measurements of the magnetic ordering, while the 
pulsed muon source at ISIS is better for measuring details of the entanglement 
between the muon and the fluorine nuclei to obtain details of the muon site.

\section{\label{sec:synth} Sample preparation}
\subsection{Synthesis}

Previous syntheses of KAgF\textsubscript{3} have utilized either a direct synthesis from KF and AgF\textsubscript{2} at elevated temperature, or a controlled thermal decomposition of a corresponding Ag(III) salt, KAgF\textsubscript{4}
\cite{Kurzydlowski2013}. The latter route is quite impractical for synthesis of a large specimen (10-20 g) which is needed for neutron studies. Therefore, we followed the former procedure while using diverse Ag precursors (AgF\textsubscript{2}, AgNO\textsubscript{3} or KAg(CN)\textsubscript{2}). A total of eleven samples were prepared. A typical procedure consisted of firing of a well ground mixture of KF and AgF\textsubscript{2} (usually with small molar excess of 1.05-1.15) in a nickel container (sometimes equipped with a teflon insert) at 300$^\circ$C for 9-10 days (with small excess of F\textsubscript{2} gas or just with argon), followed by spontaneous cooling. The purity of each specimen was scrutinized using powder x-ray diffraction utilizing a laboratory x-ray source. The samples which proved to contain some unreacted AgF\textsubscript{2} or K\textsubscript{2}AgF\textsubscript{4} layered perovskite (which both order ferromagnetically) were discarded. Only the five samples which were crystallographically pure or which contained no more than 1 percent of diamagnetic AgF were selected for further studies. All these batches were mixed together and homogenized in a prefluorinated agate mortar.

\subsection{Sample operations}
Due to the exceptionally high reactivity of Ag(II) fluorides \cite{Grochala2001}, the sample selected for neutron diffraction studies was placed inside a sealed vanadium container only shortly prior to measurements. Dry ice was used to cool the sample  during its transportation, and the container was kept permanently in liquid nitrogen in  ILL. Similarly, a small fraction of the sample (ca. 0.5-1g) from the same batch was filled into a copper container equipped with a gold O-ring, and this container was immediately chilled and transported to the relevant muon facility, where it was handled inside an argon-filled glovebox.

\begin{figure}
    \centering
    \includegraphics[width=\columnwidth]{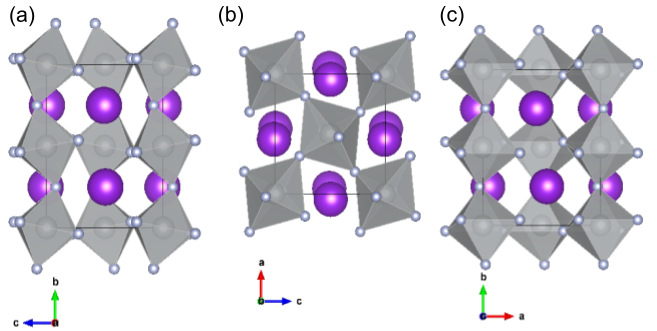}
    \caption{Panels (a), (b) and (c) show projections of the structure along the crystallographic axis ($a$,$b$ and $c$ respectively) showing octahedral rotations. Thin black lines indicate the orthorhombic unit cell.}
    \label{fig:abc}
\end{figure}

\section{Powder neutron diffraction}
\label{sec:pnd}

We performed powder neutron diffraction studies on KAgF$_3$ to determine the magnetic structure
on the powder diffractometer D20 at the Institut Laue-Langevin in Grenoble.
We choose a large wavelength of \SI{2.42}{\angstrom} and optimized the instrument for higher flux to
focus on magnetic superstructure reflections. Due to the expected small moment of Ag$^{2+}$ rather long counting was
needed and indeed the observed magnetic Bragg peaks exhibit an intensity of only 0.4\% \ compared to the strongest nuclear Bragg peak.
Data sets were collected at \SI{70}{\kelvin} ( $>T_{N1},T_{N2}$), \SI{40}{\kelvin} (between $T_{N1}$ and $T_{N2}$) and \SI{2}{\kelvin} ($<T_{N1},T_{N2}$) to cover the two magnetic transitions deduced from magnetization studies %to occur at 35 and 66\,K
\cite{Kurzydlowski2013}.
Data is available at Ref. \onlinecite{data-5-31-2635} and refinements of magnetic and nuclear structure models were performed with
the Fullprof program suite \cite{fullprof}.

Previous powder x-ray diffraction studies on  KAgF$_3$ observed a distorted perovskite structure, which is described with spacegroup $Pnma$ \cite{Kurzydlowski2013,mazej2009}. Distortions with this symmetry are very common in perovskites \cite{zhou2005}, and correspond to a rotation of the octahedron around the orthorhombic $b$ axis [Fig.~\ref{fig:abc}(b)], %which is parallel to an Ag-F bond, 
combined with tilting around nearly $a$ which is parallel
to an octahedron edge [Fig.~\ref{fig:abc}(a)]. In addition, there is an elongation of the octahedra along one of the bonds in the $a,c$ plane. This elongation rotates for neighboring sites in a $a,c$ layer. For example, the long bonds are oriented approximately along [101] in 
Fig.~\ref{fig:abc}(b), for the four octahedra in the corners, and nearly along [10$\bar1$] for the octahedron in the center. Notice that
due to the antiphase rotation of the
octahedra all elongations approach the $a$ axis forming an angle of \SI{34}{\degree} with it. 

The elongation of octahedra in KAgF$_3$
can be rationalized as originating in the Jahn-Teller distortion in the $d^9$ configuration and 
indicates an orbital order with holes alternatingly occupying $x^2-y^2$ and $z^2-y^2$ orbitals corresponding to the staggered pattern of plaquettes in Fig.~\ref{fig:struct} (notice that here long Ag-F bonds have not been drawn).
This orbital ordering determines \cite{Zhang2011} the magnetic interactions as mentioned in the Introduction.

\begin{figure}
\includegraphics[width=\columnwidth]{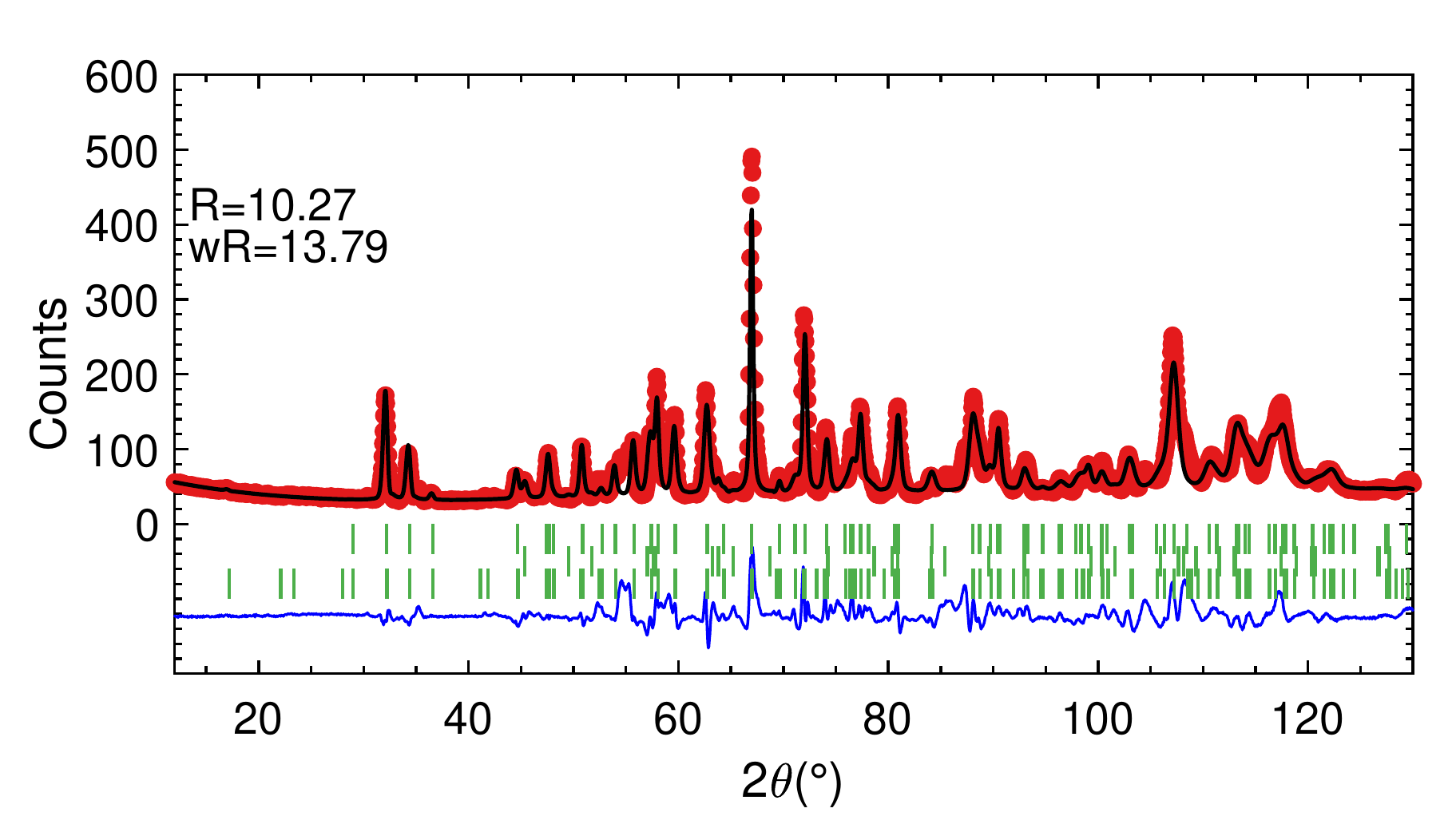}
\caption{ {Rietveld description of the data obtained on the D20 diffractometer at 2\,K.
Red symbols denote measured intensities, black dots the calculated ones, the blue line the differences, and vertical bars indicate the positions of Bragg reflections of the $Pnma$ KAgF$_3$ and AgF$_2$ phases and of the KAgF$_3$ magnetic phase.}}
\label{pattern}
\end{figure}

The D20 data taken at 2\,K can be well described with such a structure model, as shown in Fig.~\ref{pattern}.
We apply a correction for microscopic strain and include a small impurity AgF$_2$ phase, about 6\% volume fraction.
The refined lattice parameters are $a$=6.4106(6)\,\AA, $b$=8.2597(7)\,\AA, and $c$=6.0609(6)\,\AA\;  and atomic positions
are: Ag at (0,0,0), K at [0.043(2),0.25,0.483(2)], F1 at [0.479(2),0.25,0.5827(13)] and F2 at [0.3159(8),0.4630(7),0.2291(12)].

\begin{figure}
\includegraphics[width=\columnwidth]{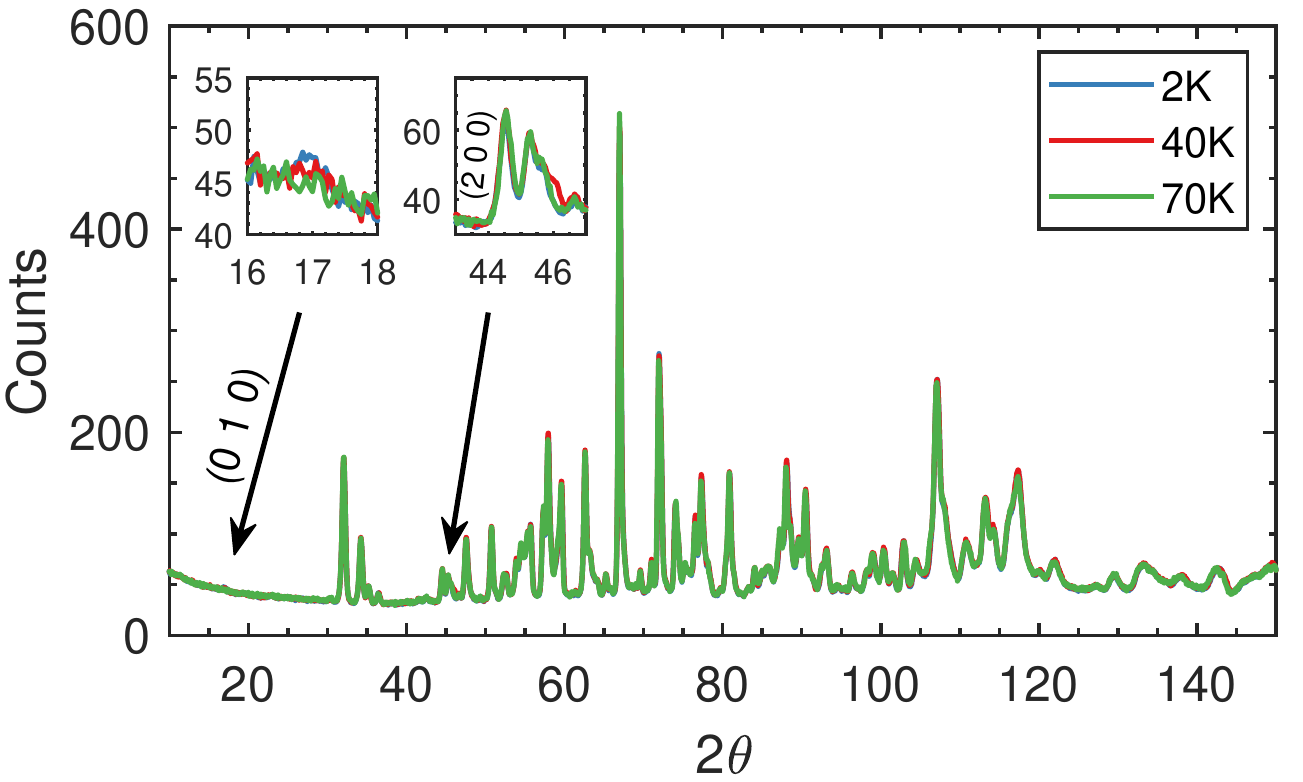}
\caption{Powder neutron diffraction data obtained on D20 at various temperatures; the two insets present the evidence for magnetic
Bragg peaks appearing in the low-temperature and intermediate magnetic phases. }
\label{D20-magnetism}
\end{figure}

The main goal of the high-flux experiments at the D20 diffractometer was to determine the magnetic structure.
We searched for magnetic Bragg reflections appearing below the N\'eel temperatures identified
by the magnetization measurements. 
The three diffraction patterns shown in Fig.~\ref{D20-magnetism} exhibit very little differences, indicating
that the crystal structure does not significantly change below 70~K. For most Bragg reflections and for the background the patterns perfectly superpose.

Due to the high statistics and the excellent reproducibility of the diffraction experiments, we are able to analyse the tiny changes in the diffraction patterns.
In the pattern at \SI{2}{\kelvin} one finds an extra peak emerging at \SI{16.9}{\degree}, which exhibits an intensity of
4$\times$10$^{-3}$ compared to the strongest nuclear Bragg peaks. This intensity can be indexed as (0,1,0) in the $Pnma$ lattice, which
is extinct in this space group. This reflection can be explained by an $A$-type magnetic order with ferromagnetic $a,c$ layers antiferromagnetically
stacked along $b$. 
Such a magnetic structure is expected for the antiferro-orbital ordering described above\cite{Zhang2011,Kurzydlowski2013}. Within an $a,c$ layer, the perpendicular
orbital arrangement with alternating occupation of the $x^2-y^2$ and $z^2-y^2$ orbitals (hole picture) implies a ferromagnetic interaction that, however,
should be weak. The strongest magnetic interaction is expected for the parallel orbital arrangement along $b$ due to strong hybridization.
This orbital and magnetic arrangement was already proposed by X. Zhang et al. \cite{Zhang2011}. Since we observe the (0,1,0) Bragg reflection as the main magnetic Bragg peak the ordered moment must point perpendicular to $b$ which is consistent with our DFT computations (See Supplementary Information) and with the muon results to be shown below. The refinement of magnetic models cannot distinguish between alignment along $a$ or $c$ direction,
but the symmetry analysis with representation theory indicates that an $A$-type order with moments along the $a$ direction permits
the occurrence of a weak ferromagnetic moment along $b$, which is not seen in the magnetization data \cite{Kurzydlowski2013}.
Therefore, only the $A$-type order with moments along $c$ is possible. 
The model yields an ordered moment of
$\mu_\mathrm{Ag}=0.47(15)$\,$\mu_B$ assuming the Pd$^{1+}$ form factor\footnote{Due to the rather small $Q$ value the form factor has little impact in this case. See the supplementary information to Ref.~\cite{Gawraczynski2019} for a comparison of the \ch{Pd1+} form factor and a DFT computation of \ch{Ag^{2+}} form factors.} and allows for a canting of moments with a $G$-type $x$ and $C$-type $y$. For the $4d$ Ag moment one may expect sizable canting but its determination is well beyond the precision of our experiment. This value of the magnetic moment  agrees well with the moment determined by the $\mu$SR experiments (as reported in the next section)
when assuming a $c$ orientation, while for an $a$ orientation the muon analysis yields a much smaller moment. This gives further support to the conclusion that the main component of the magnetic moment in KAgF$_3$ is parallel to $c$.

The (0,1,0) magnetic intensity is absent at 40\,K (see Fig.~\ref{D20-magnetism}) consistent with the lower magnetic transition observed in the
susceptibility at $T_\mathrm{N1}\approx35$~K \cite{Kurzydlowski2013}. In the data at 40\,K however, there is extra intensity at \SI{46}{\degree} that is not visible neither in the \SI{2}{\kelvin} nor in the \SI{70}{\kelvin} data. Note that the higher magnetic transition is found at \SI{66}{\kelvin} in
the magnetization. This intensity can be indexed as (1.3,1,1.3) but with the single magnetic reflection the magnetic structure cannot be determined.
Possibly this magnetic structure not only differs in the propagation vector but also in the direction of the ordered moment.
A tempting possibility would be a spiral, possibly commensurate magnetic structure at (1/3,0,1/3). Notice that in this case, the canting of adjacent moments would occur along small-$J$ bonds. Indeed, our DFT computations lead to a very small cost of the spiral (see Supplementary Information).
Thus, it is plausibly that the small energetic penalty of forming the spiral is  overcome by entropic effects. Such intermediate spiral phases are common in frustrated systems, for example in the multiferroic phase of CuO\cite{Kimura2008,Giovannetti2011a,Hellsvik2014}.  
Here, such a possibility would make \ch{KAgF3} an analogue of CuO, perhaps with multiferroic properties.  

In the next Section, we will report on the magnetic phases from the point of view of 
$\mu$SR spectroscopy. This  will provide further information on the magnetic phases.  
The $\mu$SR experiment, however, does not find evidence for magnetic order at this temperature range. This discrepancy requires further analysis.

\section{$\boldsymbol{\mu}$SR Results and Analysis}
\label{sec:musr} 
$\mu$SR spectroscopy is an excellent technique to detect 
local magnetic environments: in this technique, positive muons 
are implanted in the sample, and stop somewhere in the crystal \cite{Blundell2022}.
Hence, muons act as local magnetometers, so for a fully
quantitative analysis of $\mu$SR data it is important to 
determine the stopping site. To this aim we have performed 
\emph{ab initio} DFT+$\mu$ calculations \cite{Moeller2013, Bernardini2013}, 
which we report now.

\subsection{Muon Site Calculations}

\begin{figure}
    \centering
    \includegraphics[width=0.5\textwidth]{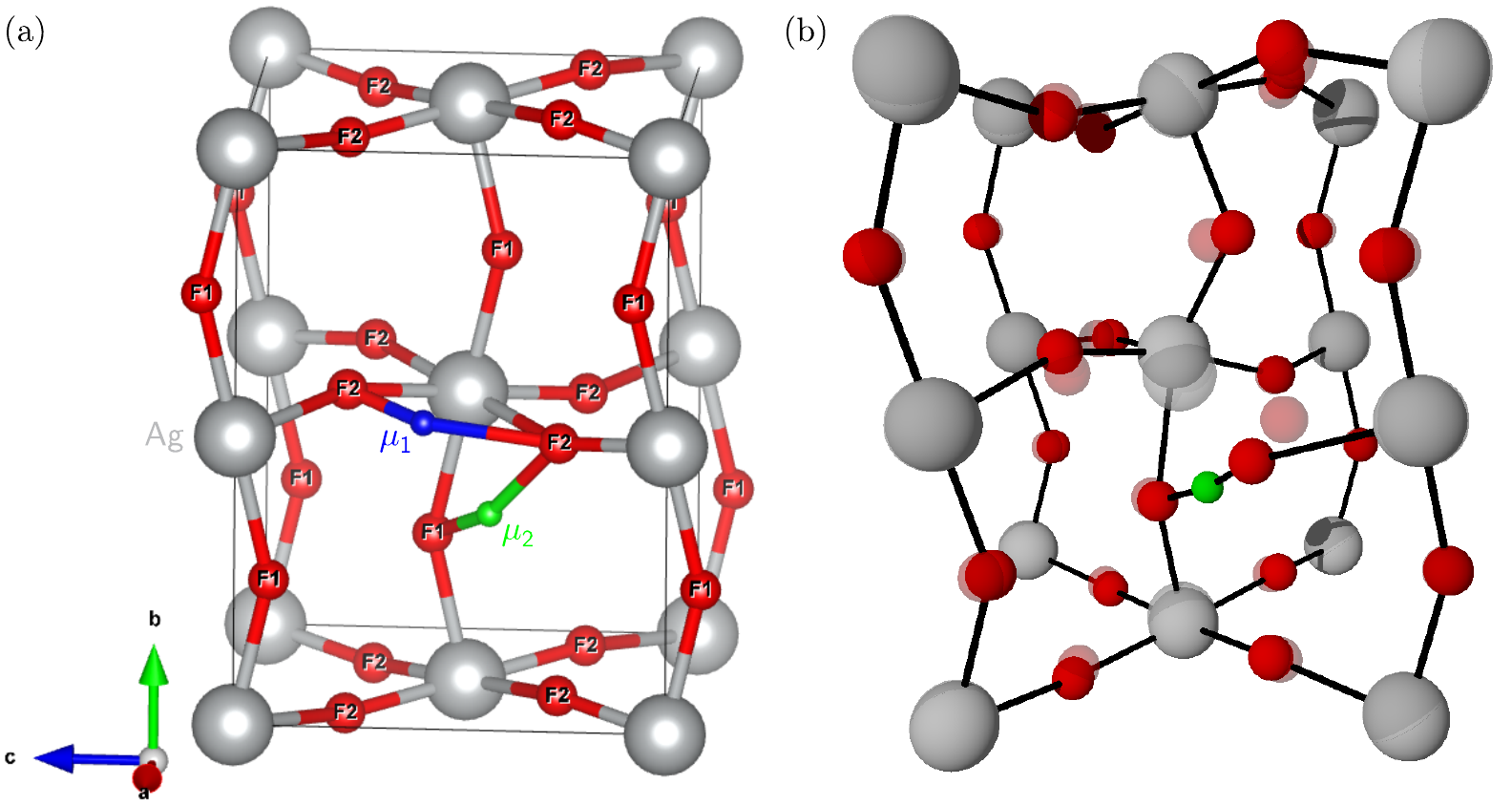}
    \caption{KAgF\textsubscript{3} muon sites. (a) shows the two candidate muon sites 
    as described in Table~\ref{tab:muonsites}, and (b) shows site 2 with its undistorted
    and distorted atoms in its unit cell as translucent and opaque spheres respectively. 
    }
    \label{fig:muonsites}
\end{figure}

The DFT+$\mu$ calculations resulted in two possible sites, which 
are shown in Table \ref{tab:muonsites}, with all other
sites discounted due to having an energy far above these. These two candidate
muon sites are depicted in  Figure~{\ref{fig:muonsites}(a)}.
As is usually the case for muons implanting in fluoride samples, DFT+$\mu$ 
predicts that the muon will draw in the two nearest fluorines towards it for both
sites. Although DFT often struggles to calculate the extent of these distortions
to a high accuracy, we estimate that for site 1 the nearest-neighbour fluorine goes from 
being  1.35~\AA\;from the muon to 1.09~\AA, and for site 2 the nearest-neighbour moves
from 1.82~\AA\;to 1.05~\AA, and we expect these bond-length estimates to be correct to within
0.1--0.2~\AA. These large lattice distortions have been 
observed in other fluorides and are attributed to the formation 
of so called ``strong" hydrogen-like bonds
\cite{Brewer1986,Noakes1993,Wilkinson2020,Wilkinson2021} analogous to the 
ones in HF$_2^-$ ions \cite{Emsley1968}. Figure~\ref{fig:muonsites}(b) shows
the extent of these lattice distortions in site 2 (site 1 has similar distortions, but 
this site is inconsistent with the data, which we will discuss later).

\begin{table}[h]
    \centering
    \begin{tabular}{c | c | c }
    \hline\hline
    Site    &   Position (fractional coordinates) & Energy (meV) \\
    \hline
    1   & $(0.7934,\:0.4950,\:0.5802)$  & 0 \\
    2   & $(0.6829,\:0.3203,\:0.4336)$  & 123 \\
    \hline\hline
    \end{tabular}
    \caption{DFT+$\mu$ results, showing the energies above the lowest energy site.}
    \label{tab:muonsites}
\end{table}
    
\subsection{$\bm{T}<\bm{T}_{\mathrm{\bf N}\bm 1}$: Measuring the collinear magnetic order} 
\label{sec:PSIdata}
\begin{figure*}
    \centering
    \includegraphics[width=0.8\textwidth]{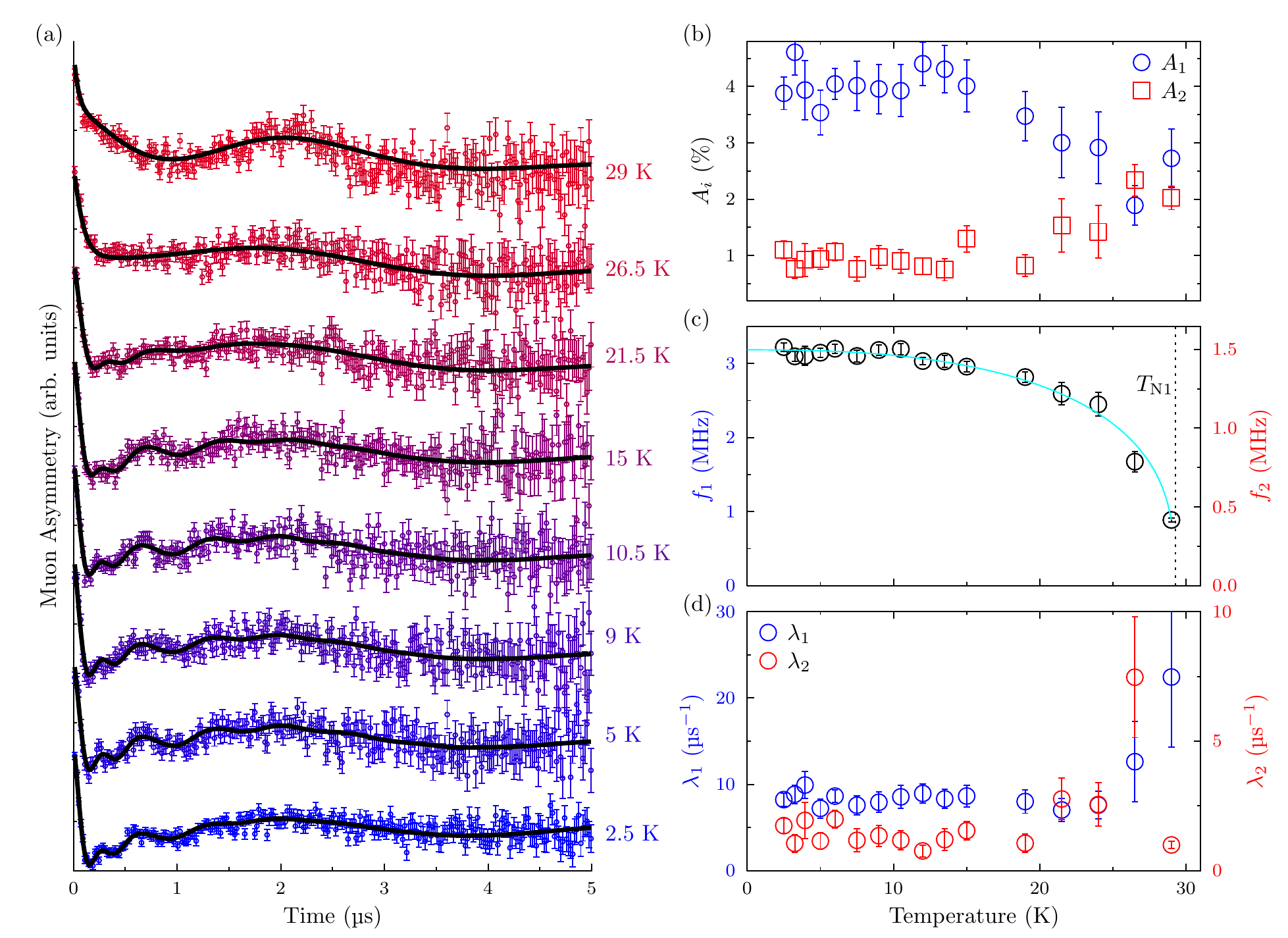}
    \caption{(a) Shows the short-time $\mu$SR data with 
             the sample holder background removed, 
             where the oscillations can be clearly observed. The black lines represent
             the fit to Equation \eqref{eq:PSIfit}, as described in the text. (b) 
             shows how the amplitudes of the two oscillating components $A_1$ and $A_2$ vary
             with temperature, (c) shows the frequency of these
             oscillations (the cyan line  shows the fit to the model as described 
             in the text), and (d) the relaxation of these
             two components.}
    \label{fig:PSIdata}
\end{figure*}

%% perhaps this title should be short-time $\mu$SR data??
The sample was contained in a Cu sample holder and placed in the GPS
spectrometer at PSI. Data from this experiment is shown in Figure \ref{fig:PSIdata}(a),
and show a clear oscillatory feature at short timescales which
is very heavily damped, and this remains present up to around 
31~K. Above 31~K, no oscillations due to magnetic ordering were observed in 
the muon asymmetry.
Some muons were found to stop in the Cu sample holder,
which creates a Kubo-Toyabe component in the muon asymmetry
\cite{Kadono1989,Luke1991}, which was modelled by the function
\begin{equation}
    A_\mathrm{KT}(t) = A_\mathrm{KT}(0)\Bigg(\frac{1}{3} +
    \frac{2}{3}(1-\Delta^2t^2)\exp\bigg(-\frac{\Delta^2t^2}{2}\bigg)\Bigg)e^{-\lambda_\mathrm{KT}t},
\end{equation}
where $\Delta$ is the standard deviation of the internal fields
at the muon site in the Cu, $A_\mathrm{KT}(0)$ is the amplitude of the asymmetry 
due to muons stopping in the sample holder, and $\lambda_\mathrm{KT}$ 
is a phenomenological parameter to model muon diffusion. 
For the $T<T_\mathrm{N1}$ data, these parameters were found to be 
$\Delta=0.1801(49)$ \textmu s$^{-1}$, $A_\mathrm{KT}(0)=11.47(6)\%$,
and $\lambda_\mathrm{KT}=0.3528(28)$ \textmu s$^{-1}$, and this function was subtracted 
from the data for all subsequent analysis.

The data, with the Cu background subtracted, were fitted with the function
\begin{equation}
    A(t) = \sum_{i=1}^3A_i\cos(2\pi f_i t - \phi_i)e^{-\lambda_it} + A_\mathrm{bg}e^{-\lambda_\mathrm{bg}t},
    \label{eq:PSIfit}
\end{equation}
where each component usually represents muons stopped in different sites in the sample,
precessing in their local magnetic fields ($B_i=2\pi f_i/\gamma_\mu$, with
$\gamma_\mu = 2\pi\times 135.5$~MHz~T$^{-1}$), with a relaxation $e^{-\lambda_it}$ due 
to field fluctuations caused  by spin dynamics and inhomogenities \cite{Blundell2022}. 
Many of these fitting parameters were found to be constant throughout the range of 
the temperatures studied. Therefore, they  were fixed to the following 
values: $\phi_1=0(10)^\circ$, $\phi_2=0.0(1)^\circ$, $A_3=0.85(16)\%$, 
$\phi_3=151(13)^\circ$, $f_3=0.277(10)$~MHz, and $\lambda_3=0.25(10)$~\textmu s$^{-1}$.
After initially being allowed to vary, it was found that the ratio between the 
frequencies $f_1$ and $f_2$ was constant at $1:0.47$, and was also fixed for the 
subsequent  analysis. The values of the parameters which were allowed to vary are 
plotted in  Figure \ref{fig:PSIdata}(b). The oscillations with amplitudes 
$A_1$ and $A_2$  are clearly due to muons stopping in the KAgF$_3$ sample, and we will
discuss the origin of these in Section \ref{sec:musranalysis}. As the $A_3$ term exhibits no 
order  parameter-like behaviour, it can be assumed to be due to a muon stopping in an 
impurity in the sample, the frequency of which is consistent with precession in a 
very small magnetic field, \emph{or} indeed it may be due the 
muon becoming entangled with fluoride impurities leading to very low amplitude 
F--$\mu$--F oscillations \footnote{See the supplemental information for a discussion 
of these states, which contains 
Refs.~\cite{Cox1987, Blundell2022, Wilkinson2020, Wilkinson2021}.}. The temperature dependence of $A_1$ and $A_2$ is very unusual 
and is a significant point we will return to later.

The $f_1$ values were fitted to the phenomenological fitting function 
\begin{equation}
    f(T) = f_0\Bigg(1-\Big(\frac{T}{T^{}_\mathrm{N1}}\Big)^\alpha\Bigg)^\beta,
\end{equation}
to model the critical behaviour of the low temperature phase. The fitting parameters
obtained were $T_\mathrm{N1}=29.26(2)$ K, $f_0=3.185(15)$ MHz, $\alpha=2.5(1)$ and 
$\beta=0.335(7)$. The critical temperature $T_\mathrm{N1}$ is slightly lower than the
35 K measured by magnetic susceptibility \cite{Kurzydlowski2013}, but this is consistent
with what was measured at ISIS (see next section and the Supplemental Information). 
The value of $\beta$ characterises the  critical behaviour of the sample, and is 
consistent with the value one would expect from a 3D Ising-like system 
(which predicts $\beta=0.326$),  though the number of data points in the critical regime is 
very low and so this should only  be taken as a crude estimate. This result suggests an 
easy axis anisotropy, which is indeed confirmed by the DFT computations reported below. 
%by inspecting the raw
%data in Figure \ref{fig:PSIwaterfall}{\bf a}, it is clear that there is no magnetic ordering 
%above 30 K, and Figure \ref{fig:PSIwaterfall}{\bf b} shows that F--$\mu$--F oscillations
%due to the \emph{very weak} fluoride nuclear moments are present at 31 K. 

\subsection{\label{sec:ISISdata}$\bm{T}>\bm{T}_{\mathrm{\bf N}\bm 1}$: Muon--fluorine entangled states} 

The sample was also placed in the MuSR spectrometer at the STFC-ISIS muon facility, 
Rutherford Appleton  Laboratory, UK. Below the transition temperature, it was 
possible to observe characteristic signs
of magnetic ordering, but it was not possible to resolve these with the time resolution of the 
spectrometer \footnote{The ISIS data for $T<T_\mathrm{N1}$ are shown in the 
Supplemental Information}.
Above the transition temperature, the characteristic oscillations expected from 
muon--fluorine entanglement were
observed \cite{Brewer1986,Noakes1993,Wilkinson2020,Wilkinson2021}.

Therefore, the muon asymmetry data for temperatures above $T_\mathrm{N1}$ were fitted to the function
\begin{equation}
    A(t) = A_\mathrm{r}P_{\mathrm{F}\mu\mathrm{F}}(r_\mathrm{nn1}, r_\mathrm{nn2}, r_\mathrm{\mu Ag}; t)
           + A_\mathrm{Cu}G_\mathrm{DKT}(\Delta, \nu; t) + A_\mathrm{bg},
    \label{eq:ISISfits}
\end{equation}
where the first term represents muons stopped in the KAgF$_3$ crystal 
and evolving due to the entanglement between itself and the surrounding
fluoride nuclei, which is calculated as described in the Supplemental 
Information. This function has the fitting parameters $r_\mathrm{nn1}$ and 
$r_\mathrm{nn2}$ to represent the distance from the muon to the closest and next-closest 
nearest neighbour fluorides respectively ($\mu_2$ to the nearest F2 and F1 in 
Figure \ref{fig:muonsites} respectively), and $r_{\mu\mathrm{Ag}}$ is the distance 
from the muon to the nearest-neighbour Ag ion. The second term represents muons stopped 
in the Cu  sample holder, and the final term represents muons stopped elsewhere 
but not undergoing any precession or relaxation which is measurable.
The Cu sample holder again produces a Kubo-Toyabe background,
which was modelled by the dynamical Kubo-Toyabe function as described by 
Kadono et al. \cite{Kadono1989} using the strong collision 
model to model the dynamics, so \footnote{Using the more rigorous strong collision 
model to model the dynamics was not for the PSI data, as the 
the dynamics only affect the data at long times and at higher temperatures.},
\begin{equation}
    G_\mathrm{DKT}(t) = g(t)e^{-\nu t} + 
        \nu \int_0^tg(\tau)e^{-\nu \tau}G_\mathrm{DKT}(t-\tau)\mathrm{d}\tau,
\end{equation}
with
\begin{equation}
    g(t) = \frac{1}{3} + \frac{2}{3}(1-\Delta^2t^2)\exp\bigg(-\frac{\Delta^2t^2}{2}\bigg).
\end{equation}

We found that calculating $P_{\mathrm{F}\mu\mathrm{F}}(t)$ for both the sites 
found with DFT+$\mu$ in Table \ref{tab:muonsites} produced a very similar 
result, meaning that the two muon sites were indistinguishable with the ISIS 
data. However, the careful analysis of the dipole fields of both sites
and the comparison of this to the PSI data (as discussed in 
Section \ref{sec:musranalysis}), shows that the only site which is realised by the muon
is site 2. 

\begin{figure*}
    \centering
    \includegraphics[width=.7\textwidth]{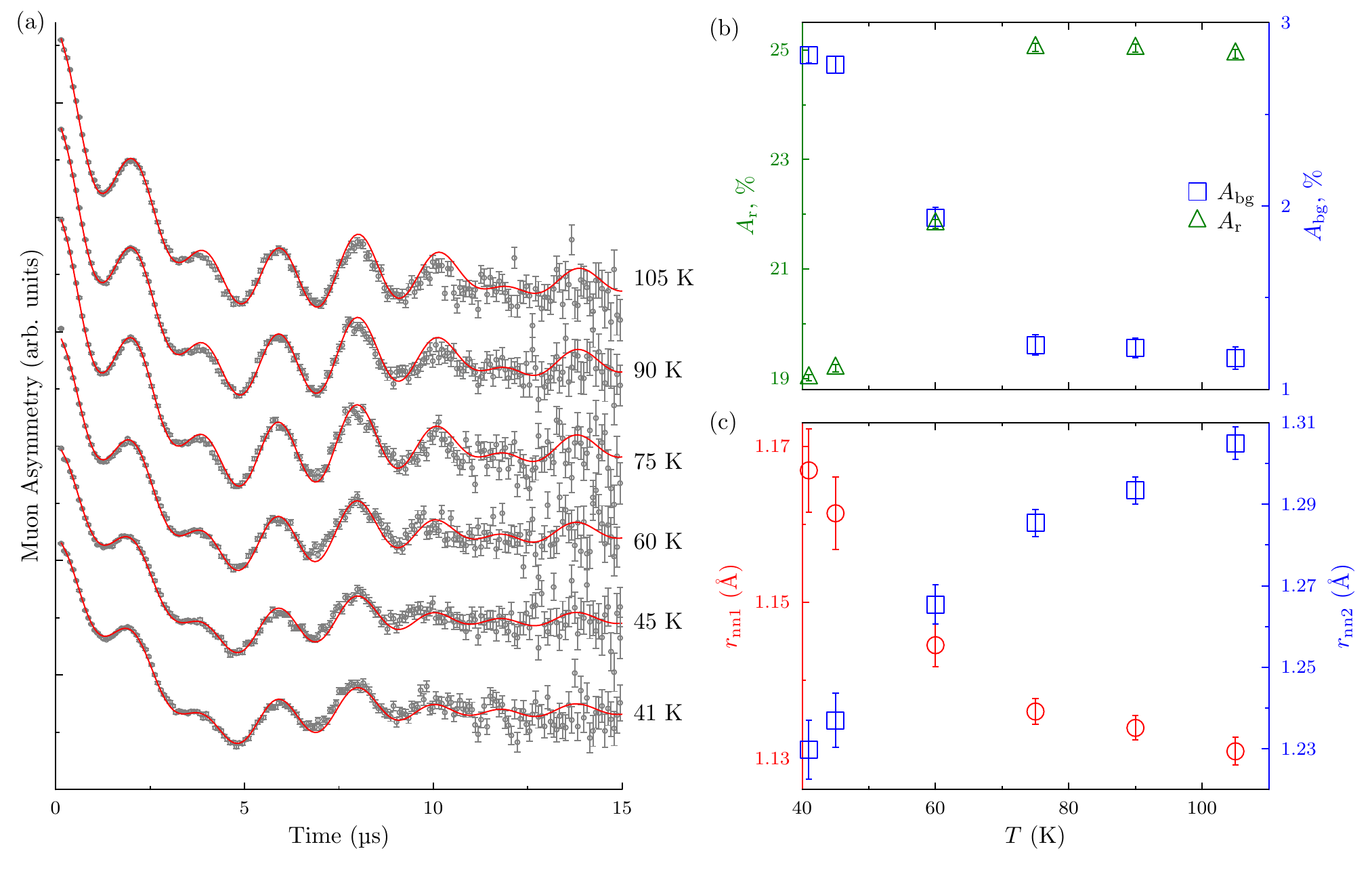}
    \caption{{\bf Fits to the ISIS data.} {\bf a} shows the fits to the data using the model described 
             in  the text.
             {\bf b} shows the variation of the relaxing asymmetry $A_\mathrm{r}$ and 
             background asymmetry $A_\mathrm{bg}$ with temperature $T$, and {\bf c} shows how the
             $\mu$--F distances $r_\mathrm{nn1}$ and $r_\mathrm{nn2}$ vary. }
    \label{fig:ISISparameters}
\end{figure*}

The fits to Equation \eqref{eq:ISISfits} are shown in Figure \ref{fig:ISISparameters}{\bf a}
and the fitting parameters obtained are plotted in Figure \ref{fig:ISISparameters}{\bf b} and {\bf c}.
Figure~\ref{fig:ISISparameters}{\bf a} demonstrates the impressive agreement between the 
model and the data up to 105~K (muon diffusion occurs at very high temperatures and this is
outside the scope of our model). The
change in $A_\mathrm{bg}$ and $A_\mathrm{r}$ is somewhat unusual, suggesting some sort of 
phase transition between 50 and 70 K, perhaps related to the 
susceptibility anomaly \cite{Kurzydlowski2013} at $T_{N2}$ and giving further support for 
the existence of an intermediate magnetic phase. Unfortunately, there is insufficient 
muon data in this temperature range to be able
to determine the cause of this. Yet, it is important to note that as 
\mbox{F--$\mu$--F} states are highly 
suppressed by magnetic fields (since the field at the muon 
site due to the ordering of the electronic moments 
tends to be much larger than that due to the nuclear moments). 

The fits also show that the F--$\mu$--F complex changes slightly with temperature, with 
a change in  bond length of around 0.08 \AA\; across the whole temperature range. 
 This is likely to be due to 
the way in which the entire structure of the crystal changes with temperature, 
as has been reported previously \cite{Kurzydlowski2013}. The DFT+$\mu$ results show 
that the two nearest neighbour fluorines are  $r^\mathrm{DFT}_\mathrm{nn1}=1.05$~\AA and 
$r^\mathrm{DFT}_\mathrm{nn2}=1.30$~\AA, which are close to the values measured here. Additionally, 
the $\mu$--Ag  distance $r_{\mu\mathrm{Ag}}$ varied very little with temperature, and was 
found to be $2.34(6)$ \AA, very close to the value of  2.35 \AA\; calculated with DFT. 

\subsection{Calculation of the Ag magnetic moment with $\boldsymbol{\mu}$SR}
\label{sec:musranalysis}

A muon implanted in a sample precessses in its local magnetic field due to the Zeeman 
interaction. Assuming that the only origin of this field in KAgF$_3$ is due to the dipolar 
field of the surrounding magnetic moments $B_\mathrm{dip}$, the local field of the muon 
can be calculated as 
\begin{equation}
\mathbf{B}_\mathrm{dip} = \sum_i \frac{\mu_0}{4\pi r_i^3}
\big(3(\boldsymbol{\mu}_i\cdot\hat{\mathbf{r}}_i)\hat{\mathbf{r}}_i - \boldsymbol{\mu}_i\big),
\label{eq:dipolefields}
\end{equation}
where the sum is over the surrounding Ag ions with a magnetic moment 
$\boldsymbol{\mu}_i$ at a distance $r_i$ from the muon. This sum is taken over a large 
Lorentz sphere of radius 150 \AA, and as the material is antiferromagnetic the lack of 
bulk magnetization means no additional term is required to account for
moments beyond this.

One can then use Equation \eqref{eq:dipolefields} to calculate the dipole fields at both the 
muon sites predicted by DFT, assuming ferromagnetically ordered moments in the 
$\mathbf{a}\mhyphen\mathbf{c}$ plane and antiferromagnetic along $\mathbf{b}$, 
with all moments constrained to point in the $\mathbf{c}$ direction, as predicted by our DFT calculations of the magnetic structure reported in the Supplemental Information.
From this, assuming 
the moment of the Ag ions $\mu_\mathrm{Ag}=0.55\mu_\mathrm{B}$, the field at the muon site would 
be 643 Gauss for site 1, and 260 Gauss for site 2, leading to muon precession frequencies of 
around 8.7 MHz and 3.5 MHz respectively. In accord with the analysis of 
Sec.~\ref{sec:PSIdata}, the precession frequency of site 2 is significantly
closer to that measured, and therefore we will assume that this is the correct muon 
site for the rest  of the analysis. As the distance between the muon and the nearest Ag 
ion was found to be 2.33(6) \AA\,by fitting the ISIS data, the moment on the Ag ion 
can be determined. Again using the dipole field calculations, 
the moment on the Ag ions is calculated as $\mu_\mathrm{Ag}=0.489(2)\mu_\mathrm{B}$, 
in excellent agreement to the value determined with neutrons (Sec.~\ref{sec:pnd}) and slightly smaller
than the value of $0.549\mu_\mathrm{B}$ predicted by our DFT results (see the Supplemental 
Information). Such an overestimation is not surprising, as is common\cite{Hirsch1989,Lorenzana2005} 
to other mean-field like approaches that neglect transverse quantum fluctuations. 

Other valid potential magnetic structures for this compound have the moments aligned 
along the $\mathbf{a}$ direction, and in any other direction along 
the $\mathbf{a}\mhyphen\mathbf{c}$ plane.  If the spins are aligned along 
$\mathbf{a}$, following the same calculation as above results in an
Ag electronic moment of $0.196(1)\mu_\mathrm{B}$, which is significantly smaller than the moment 
calculated both from the neutron data and by the DFT calculations, and is therefore an unlikely 
magnetic structure. If the moments are aligned along an intermediate direction in 
the $\mathbf{a}\mhyphen\mathbf{c}$ plane, one would expect the muon data to have two 
frequencies with  an equal magnitude (owing to the site symmetry), which is not realised 
in the data (although it  is important to note that the measured muon precession at 
this frequency relaxes very  quickly, so moments pointing in an intermediate 
direction cannot conclusively be ruled out). 

\subsection{Origin of the two frequencies in the PSI data}

\begin{figure*}
    \centering
    \includegraphics[width=\textwidth]{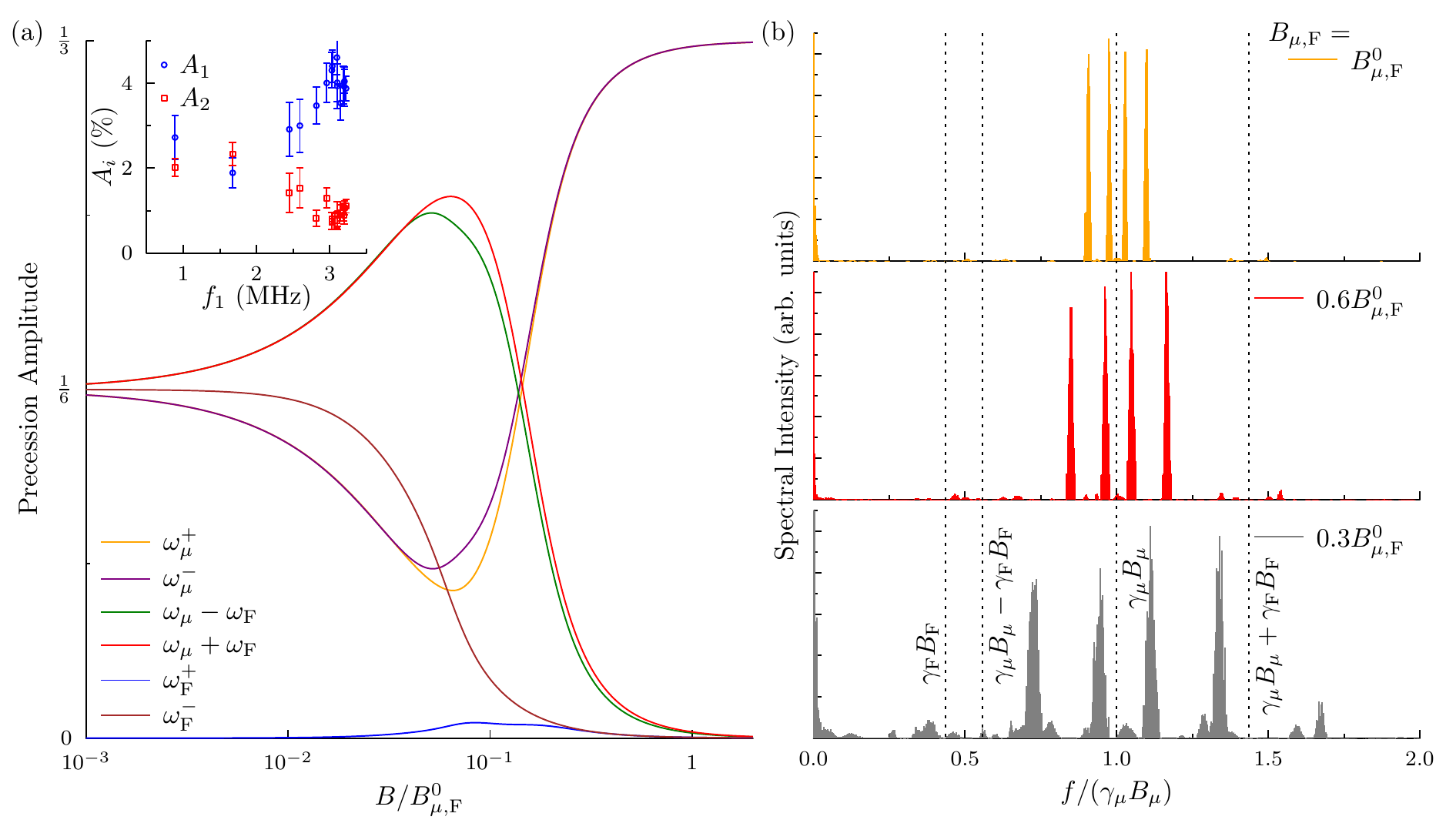}
    \caption{F--$\mu$ states in a magnetic field:
             (a) shows how the amplitudes of the components of the 
             muon precession signal vary as the field on the muon 
             and fluorine sites are scaled by a factor  
             $B/B_{\mu, \mathrm{F}}^0$ (the key shows the value of these
             precession frequencies as $B_\mathrm\rightarrow\infty$). The 
             inset shows how the amplitudes $A_1$ and $A_2$ of the two 
             measured components vary with the muon precession frequency,
             showing how at low fields $A_1$ and $A_2$ become closer 
             in value, in support of this model.
             (b) shows the Fourier transform of the system of the 
             muon coupled to its nearest- and next-nearest neighbour 
             spins, for the field at the muon and fluorine sites scaled by 
             various amounts, as shown in the legend for each of the subplots.}
    \label{fig:FmuB}
\end{figure*}

As discussed in Section \ref{sec:PSIdata}, the PSI data show two oscillations with 
frequencies $f_1$ and $f_2$ which follow the same order parameter.  
The two frequencies could be associated with two distinct muon sites,
but the temperature-dependent nature of the amplitudes $A_1$ and $A_2$ shown in 
Figure~\ref{fig:PSIdata}(b), together with the single site assumed for fitting 
the F--$\mu$--F oscillations (which works so well, as shown in 
Figure~\ref{fig:ISISparameters}(a) point  instead to a different interpretation.
%Although one
%may naively assign both these frequencies to muons stopping in multiple sites, 
%it is important to note that the relative amplitude of both these oscillations 
%is {\emph vastly} different for $T\ll T_{N1}$ ($A_2$, the amplitude of the slower oscillation,
%is much smaller than $A_1$), the ISIS data does not support multiple muon 
%sites (although it struggles to distinguish between both sites from DFT, other
%sites, especially those with a distance from the Ag moments that would 
%produce a field of the measured frequency would likely be visible \cite{Wilkinson2021}).
%
We propose that the origin of the second oscillation is due to the effect of the 
nearest-neighbour  fluorine nucleus precessing in its own local magnetic field resulting 
from the ordered  Ag$^{2+}$ electronic moments, which then affects the state of the muon. 
One can picture this by  considering both the muon and fluorine to be classical dipoles 
in a magnetic field. Before the muon implantation, the fluorine nuclear spin is in a fully 
mixed state. But once the polarised muon is  implanted and the lattice distorted, the 
fluorine becomes slightly polarised by the muon, and the field at the fluorine 
is  modified due to the strong lattice distortion. Hence, the fluorine 
and muon \emph{both} undergo Larmor precession in the 
field produced by the ordered Ag$^{2+}$ moments,  but the dipole-dipole interaction between 
them creates  a field at the muon site that depends 
on the orientation of the nuclear spin of the fluorine with respect to the local field direction 
at the fluorine site.  The field at the nearest-neighbour fluorine site (assuming the 
magnetic structure  of KAgF$_3$ described above) is calculated as 347~Gauss, which is 
larger than that of the muon (235 Gauss), and points in a different direction.

In order to model the two frequencies, one can calculate the expected amplitudes and 
frequencies  of the muon polarization using the Hamiltonian in the Supplemental information, for the case of a system with a muon and 
one fluorine (the other fluorine is further away, so does not have such a large effect). 
This produces the precession frequencies (for a large magnetic field) of
\begin{align*}
\omega_\mu^\pm &= \gamma_\mu B_\mu \pm \delta_\mu \\
\omega_\mathrm{F}^\pm &= \gamma_\mathrm{F} B_\mathrm{F} \pm \delta_\mathrm{F} \\
\omega_\mathrm{F} \pm \omega_\mu &= \gamma_\mathrm{F} B_\mathrm{F} \pm \gamma_\mu B_\mu,
\end{align*}
where $\delta_\mu$ and $\delta_\mathrm{F}$ are the splittings due to the dipole-dipole coupling
between the muon and the fluorine, which depends on the geometry of the F--$\mu$ bond 
with respect to the magnetic fields, and is around $0.2003\omega_\mathrm{D}$ for this case 
($\omega_\mathrm{D}=\frac{\mu_0 \gamma_\mu \gamma_\mathrm{F}}{4 \pi \hbar |\mathbf{r}|^3}
= 1.507$ Mrad s$^{-1}$ from the ISIS data for this geometry). The amplitude of the 
signals corresponding to these states is  plotted in Figure \ref{fig:FmuB}(a), for the field 
at the muon and fluorine sites, scaled by  various amounts. This shows that, when the 
fields at the muon and fluorine sites are very large,  the only precession frequencies 
observable are due to the muon precessing at its Larmor frequency. 
However, for the relatively small fields seen in KAgF$_3$, the figure shows that the 
$\omega_\mathrm{F}^\pm$ and $\omega_\mathrm{F}\pm\omega_\mu$ terms increase as the 
magnetic fields decrease, which occurs when the temperature approaches the transition.

The inset to Figure \ref{fig:FmuB}(a) shows how the amplitudes of the two components 
$A_1$ and $A_2$  vary with the precession frequency $f_1$. This shows a clear trend of 
$A_1$ decreasing as $A_2$  increases, which is consistent with this model, despite 
the effect being more pronounced in the data as one would expect from the theory. 
Expanding this model to consider also the next-nearest neighbour
interactions, the amplitudes of these oscillations change slightly, as shown in 
Figure~\ref{fig:FmuB}(b). This shows the Fourier components of the expected muon 
polarization, showing peaks close to many of the  frequencies of the F--$\mu$ case
(but many of these are broadened out by the next-nearest neighbour dipole-dipole 
interactions making them very small).
Decreasing the field further, it can be seen that the peaks just below around 50~\% of the 
muon Larmor frequency start to emerge, which is broadly consistent with the $f_1:f_2$ 
ratio found previously. 

\section{Conclusions}
\label{sec:conc}

\begin{table}[h]
    \centering
    \begin{tabular}{c | c | c | c | c| c}
    \hline\hline
    \multirow{1}{*}{Compound} & \multirow{1}{*}{T\textsubscript{N}} & \multirow{1}{*}{J\textsubscript{1D}} & \multirow{1}{*}{J\textsubscript{perp}} & \multirow{1}{*}{T\textsubscript{N}}/J\textsubscript{1D} &   \multirow{1}{*}{-J\textsubscript{perp}}/J\textsubscript{1D} \\
    \hline
    KCuF\textsubscript{3}   & 39  & 406 & -21 & 9.6 x $10^{-2}$ & 5.2 x $10^{-2}$ \\
    KAgF\textsubscript{3}   & 29.3  & 1160& -13.5 & 2.5 x $10^{-2}$ & 1.2 x $10^{-2}$ \\
    \hline\hline
    \end{tabular}
    \caption{The Neel point, intra-chain and inter-chain magnetic superexchange constants (in units of Kelvin), reduced Neel temperature and magnetic anisotropy for KCuF\textsubscript{3} and KAgF\textsubscript{3}). Experimental values of N\'eel temperature and exchange constants for  KAgF\textsubscript{3} are from Ref.~\cite{Kurzydlowski2013}. For KCuF\textsubscript{3} values are  from    Ref.~\cite{DePinhoRibeiroMoreira1999} and references therein. }
    \label{tab:AgCucompare}
\end{table}

The combined powder neutron diffraction and muon spin rotation studies as well as theoretical DFT calculations on KAgF\textsubscript{3} allowed us to unequivocally determine the magnetic ground state of this compound as
an ordered A-type antiferromagnet with a N\'eel temperature $T_{N1}=29$~K. This is close to the temperature of 35~K, where an anomaly was found in the susceptibility data \cite{Kurzydlowski2013}.  In Ref.~\cite{Kurzydlowski2013}, another anomaly was reported at $T_{N2}\approx 66$K, so it is natural to ask if there is an intermediate magnetically ordered phase between $T_{\rm N1}$ and $T_{\rm N2}$. Neutron diffraction experiments provide evidence for an incommensurate phase in this temperature region, however the signal is rather weak and insufficient to determine the magnetic structure. A strong temperature dependence of a background contribution in the analysis of the muon data suggest some kind of phase transition in the same temperature range. On the other hand, there is no evidence of static moments, even with an incommensurate arrangement, as would be suggested by the neutrons. Therefore, the possibility of an intermediate magnetic phase between the ground state and the disordered paramagnet, analogous to CuO \cite{Kimura2008}, remains an open problem. 
Another interesting open problem is the origin of the structural transitions near $T=235$~K  and the relationship between this with the magnetic ordering, if any. A detailed structural study is under way to solve this issue.

It is interesting to compare the present results with 
\ch{KCuF3}. Following Yasuda et al.\cite{Yasuda2005}, we may now 
determine the value of J\textsubscript{perp}, which turns out to 
amount to $-13.5$~K (as shown in Table~\ref{tab:AgCucompare}). From our DFT computations we estimate a value of $J_{1D}=173$ meV (~2000 K) and taking an effective $|J_{\rm perp}|=|Jac+(Ja+Jc)/2|\approx 4.25 $ (~50 K)  meV (obtained by averaging over
calculated interchain exchange interactions, see Supplementary Information) we obtain 
 $|J_{\rm perp}/J_{1D}|=2.5\times 10^{-2}$, of the same order of magnitude as the experimental one. The theoretical $J_{1D}  $  appears somewhat larger than the experimental result and previous DFT studies\cite{Zhang2011,Kurzydlowski2013}, probably because of details on the functional used. We notice, however, that even if we plug the calculated $J_{1D}$ in the formula of Yasuda et al.\cite{Yasuda2005}, we get $J_{\rm perp}=-12.8~$K, with a marginal impact on the ratio $|J_{\rm perp}/J_{1D}|=1.1\times 10^{-2}$.  Therefore,
KAgF\textsubscript{3} seems to exhibit a ca. four times larger bond anisotropy in inter- and intra-chain exchange interactions than its Cu analogue, which  is manifested in 
a smaller T\textsubscript{N} value for the former, and 
the  J\textsubscript{1D} being at least 3-times larger for the Ag 
than for the Cu compound.

\section{Acknowledgements}
\label{sec:ackn}

Research was carried out with the use of CePT infrastructure financed by the European Union - the European Regional Development Fund within the Operational Programme “Innovative economy” for 2007-2013 (No. POIG.02.02.00-14-024/08-00). The Polish authors are grateful to NCN for support (Maestro, No. 2017/26/A/ST5/00570). The Italian authors acknowledge financial support from the Italian MIUR through Projects No. PRIN 2017Z8TS5B and 20207ZXT4Z. The Slovenian authors acknowledge the financial support of the Slovenian Research Agency (research core funding No. P1-0045; Inorganic Chemistry and Technology).
The German authors acknowledge funding by the Deutsche Forschungsgemeinschaft (DFG, German Research Foundation) Project No. 277146847 - CRC 1238, Project B04.
W.G. is grateful to the Interdisciplinary Center for Mathematical and Computational Modelling, University of Warsaw, for the availability of high performance computing resources (okeanos, topola) within the projects No. G29-3 and GA83-34. 
The muon data for $T<T_\mathrm{N1}$ were taken at the Swiss Muon Source, PSI, Switzerland,
and J. M. W. and S. J. B. would like to thank Chennan Wang for running the experiment. 
The muon data for $T>T_\mathrm{N1}$ were taken at the ISIS Neutron and Muon Source, UK, 
and we would like to thank Francis Pratt for his assistance \cite{ISISdata}. The DFT+$\mu$ calculations 
were done both on the Redwood cluster at the University of Oxford with the assistance of 
Jonathan Patterson, and also using the STFC's SCARF cluster.


%\section{References}
%\label{sec:ref}

\bibliographystyle{apsrev4-2}

\bibliography{bibliography}% Produces the bibliography via BibTeX.

\end{document}